\documentclass[aps,prd,showpacs,twocolumn]{revtex4}

\newcommand{\be}{\begin{eqnarray}}
\newcommand{\ee}{\end{eqnarray}}
\newcommand{\nn}{\nonumber\\}

\newcommand{\sfrac}[2]{{\textstyle\frac{#1}{#2}}}
\newcommand{\sint}{{\textstyle\int}}

\newcommand{\Det}{\mathrm{Det}}
\newcommand{\I}{\mathrm{i}}

\newcommand{\ssh}{\hskip -0.7mm\not\hskip -0.7mm}
\newcommand{\SSH}{\hskip -1mm\not\hskip -1mm}

\usepackage{bbm}

\newcommand{\vs}{\vskip 8mm}

\begin{document}

\title{Concerning gauge field fluctuations around classical configurations}
\author{Dennis D.~Dietrich}
\affiliation{Center for Particle Physics Phenomenology, University of Southern Denmark, Odense, Denmark}
\date{June 8, 2009}

\begin{abstract}

We treat the fluctuations of non-Abelian gauge fields around a classical
configuration by means of a transformation from the Yang--Mills
gauge field to a homogeneously transforming field variable. 
We use the formalism to compute the effective action induced by these fluctuations in a
static background without Wu--Yang ambiguity.

\end{abstract}


\maketitle


The response of quanta to classical gauge fields is a fundamental issue of
continuing phenomenological and theoretical interest in Abelian and
non-Abelian field theories. The canonical example is the production of
electron-positron pairs in strong photon fields \cite{ConstantFieldHistory}
which among other effects is about to be tested with ultra strong light
sources \cite{ELI}. It is one thing if the aforementioned quanta are matter
particles like fermions or scalars and another if they are fluctuations of
the gauge field around its expectation value: As the entire gauge field does
not transform homogeneously under gauge transformations, at variance with
the matter fields, keeping track of gauge invariance when handling the
fluctuations has an extra twist to it. At leading order this is
only important for self-interacting, i.e., non-Abelian fields. One way is to
use the Faddeev--Popov approach, especially in conjunction with the 
background field method \cite{BFM}. Here we will use another approach
based on a transformation from the vector gauge field $A^a_\mu$ to an
antisymmetric tensor variable $B^a_{\mu\nu}$, passing via a first-order formulation. 
$B^a_{\mu\nu}$ 
transforms homogeneously under gauge transformations. We derive the general formalism and then
compute the effective Lagrangian for an example. Before we delve into the non-Abelian case we 
first take a look at the Abelian, where $B_{\mu\nu}$ is even gauge invariant.


\vs

{\it Abelian.} 
The partition function of quantum electrodynamics coupled to an external
source $J^\mu$ is given by,
\be
Z
&=&
\int[dA][d\psi][d\bar{\psi}]\exp\{i\sint
d^4x[-\sfrac{1}{4}F_{\mu\nu}F^{\mu\nu}+\nn&&+\bar{\psi}(i\SSH D-m)\psi-A_\mu
J^\mu]\} ,
\label{gf:qed}
\ee
where $F_{\mu\nu}=\partial_\mu A_\nu-\partial_\nu A_\mu$ represents the field
tensor, $A_\mu$ the gauge field, $D_\mu=\partial_\mu-igA_\mu$ the covariant
derivative, $g$ the coupling constant, $m$ the fermion mass, and
$\psi/\bar{\psi}$ the fermion fields. Integrating out $A_\mu$ leads to,
\be
Z
&\cong&
\int[d\psi][d\bar{\psi}]\exp\{i\sint
d^4x[\bar{\psi}(i\ssh\partial-m)\psi]\}\times\nn&&\times\exp\{i\sint
d^4xd^4y[-\sfrac{1}{2} 
(g\bar{\psi}\gamma_\mu\psi-J_\mu)(x)\times\nn&&~\times\Gamma^{\mu\nu}(x-y)
(g\bar{\psi}\gamma_\nu\psi-J_\nu)(y)]\} ,
\label{gf:qed:noa}
\ee
where $\Gamma^{\mu\nu}(x-y)$ stands for the photon propagator in some
gauge. $\cong$ indicates that the normalisation was changed in that step.
The terms in the second exponential describe single-photon exchange
and couple the fermions to the background.
The stationarity condition for $\bar{\psi}$ yields the Dirac equation for
$\psi$ in the background $\mathcal{A}_\mu$,
\mbox{$
(i\SSH\mathcal{D}-m)\psi=0
$},
where $\mathcal{D}_\mu$ stands for the covariant derivative on
$\mathcal{A}_\mu$. The latter is the solution of the
stationarity condition for $A_\mu$ in the action in Eq.~(\ref{gf:qed}),
\be
\partial_\mu\mathcal{F}^{\mu\nu}=J^\nu-g\bar{\psi}\gamma^\nu\psi ,
\label{maxwell}
\ee
where $\mathcal{F}_{\mu\nu}$ is the field tensor on the classical solution.

We replace the gauge field by an antisymmetric tensor field $B_{\mu\nu}$. This is 
achieved by multiplying Eq.~(\ref{gf:qed}) by a Gaussian integral over
$B_{\mu\nu}$, followed by a shift of $B_{\mu\nu}$ by the dual field tensor $\tilde
F^{\mu\nu}=\sfrac{1}{2}\epsilon^{\mu\nu\kappa\lambda}F_{\kappa\lambda}$,
\be
Z&\cong&\int[dA][dB][d\psi][d\bar{\psi}]\exp\{i\sint
d^4x[-\sfrac{1}{4}B_{\mu\nu}B^{\mu\nu}-\nn&&-\sfrac{1}{2}B_{\mu\nu}\tilde F^{\mu\nu}+\bar{\psi}(i\SSH
D-m)\psi-A_\mu J^\mu]\} .
\label{gf:bf}
\ee
Postulating
the gauge invariance of the $BF$ term requires a gauge invariant $B_{\mu\nu}$.
The stationarity conditions for $A_\mu$ and $B_{\mu\nu}$,
\be
\partial_\mu\tilde{\mathcal{B}}^{\mu\nu}=J^\nu-g\bar{\psi}\gamma^\nu\psi
~~~\mathrm{and}~~~
\mathcal{B}^{\mu\nu}=-\tilde{\mathcal{F}}^{\mu\nu} ,
\label{cleoms:bf}
\ee
combine into Eq.~(\ref{maxwell}), where $\mathcal{B}_{\mu\nu}$ is the classical value
of $B_{\mu\nu}$. 
Integrating out $A_\mu$ in Eq.~(\ref{gf:bf}) yields,
\be
Z&\cong&\int[dB][d\psi][d\bar{\psi}]\delta(\partial_\mu\tilde
B^{\mu\nu}-J^\nu+g\bar{\psi}\gamma^\nu\psi)\times\nn&&\times\exp\{i\sint
d^4x[-\sfrac{1}{4}B_{\mu\nu}B^{\mu\nu}+\bar{\psi}(i\ssh\partial-m)\psi]\} ,
\label{gf:delta}
\ee
which required no gauge fixing and yields a local result. The first of 
Eqs.~(\ref{cleoms:bf}) is now strictly enforced; it does not merely give the 
most probable configuration, but the only allowed configuration. This 
constraint can be used to eliminate
$B_{\mu\nu}$ from the partition function, such that
\be
B_{\mu\nu}=\sfrac{1}{2}\epsilon_{\kappa\mu\nu\lambda}\int
d^4yS^\kappa(x-y)(J^\lambda-g\bar{\psi}\gamma^\lambda\psi)(y) ,
\label{B}
\ee
where in momentum space $\mathring{S}^\kappa(p)=ip^\kappa p^{-2}$ with an
appropriate pole prescription. 
Replacing $B_{\mu\nu}$ in the exponent of
Eq.~(\ref{gf:delta}) by Eq.~(\ref{B}) leads to Eq.~(\ref{gf:qed:noa}) with 
$\Gamma^{\mu\nu}$ transverse which corresponds to Landau gauge.

The $\delta$ constraint in Eq.~(\ref{gf:delta}) eliminated $B_{\mu\nu}$,
resulting in a theory of interacting fermions, reproducing 
Eq.~(\ref{gf:qed:noa}), but without fixing a gauge.


\vs

{\it Non-Abelian.}
Consider the generating functional for a Yang--Mills field $A^a_\mu$ coupled 
to an external source $J^a_\mu$, 
\be
Z=\int[dA]\exp[\I\sint d^4x(
-\sfrac{1}{4}F^a_{\mu\nu}F^{a\mu\nu}-A_\mu^aJ^{a\mu}
)],
\ee
where
$
F^a_{\mu\nu}
=
\partial_\mu A^a_\nu-\partial_\nu A^a_\mu+gf^{abc}A^b_\mu A^c_\nu
$
stands for the field tensor, $g$ for the coupling constant, and $f^{abc}$
for the antisymmetric structure constant of the gauge group. The
corresponding classical equations of motion read,
\be
\mathcal{D}_\mu^{ab}\mathcal{F}^{b\mu\nu}=J^{a\nu},
\label{YMEq}
\ee
where $\mathcal{D}_\mu^{ab}$ represents the covariant derivative 
$D_\mu^{ab}=\delta^{ab}\partial_\mu+gf^{acb}A_\mu^c$ and
$\mathcal{F}^a_{\mu\nu}$ the field tensor $F^a_{\mu\nu}$ both on the classical
solution $\mathcal{A}^a_\mu$ for the gauge field. We will now split the
gauge field according to
$
A_\mu^a=\mathcal{A}^a_\mu+a^a_\mu ,
$
and introduce an antisymmetric tensor field in the same way as in the
previous section. Doing so yields $Z$ in the so-called first-order formalism \cite{Deser:1976iy},
\be
Z\cong\int[da][dB]\exp[\I\sint d^4x(
-\sfrac{1}{4}B^a_{\mu\nu}B^{a\mu\nu}
-\nn-\sfrac{1}{2}\tilde B^a_{\mu\nu}F^{a\mu\nu}
-A_\mu^aJ^{a\mu}
)].
\label{YMBF}
\ee
A homogeneously transforming $B_{\mu\nu}\rightarrow UB_{\mu\nu}U^\dagger$, 
leads to a gauge invariant action, for $J^a_\mu\equiv0$. Integrating out 
$a^a_\mu$ 
we find,
\be
Z&\cong&\int[dB]\Det^{-\sfrac{1}{2}}\mathbbm{B}\exp\{\I\sint d^4x[
\label{gf:BF}\\
&&-\sfrac{1}{4}B^a_{\mu\nu}B^{a\mu\nu}
-\sfrac{1}{2}\tilde B^a_{\mu\nu}\mathcal{F}^{a\mu\nu}
-\mathcal{A}^a_\mu J^{a\mu}
+\nn
&&+\sfrac{1}{2}
(\mathcal{D}^{ac}_\kappa\tilde B^{c\kappa\mu}-J^{a\mu})
(\mathbbm{B}^{-1})^{ab}_{\mu\nu}
(\mathcal{D}^{bd}_\lambda\tilde B^{d\lambda\nu}-J^{b\nu})
]\} ,
\nonumber
\ee
where
$
\mathbbm{B}^{bc}_{\mu\nu}=g\tilde B_{\mu\nu}^a f^{abc}
$
and
\be
\Det^{-\sfrac{1}{2}}\mathbbm{B}
&\cong&
\int[d\zeta]\exp[-\sfrac{\I}{2}\sint d^4x
(\zeta^{a\mu}\mathbbm{B}^{ab}_{\mu\nu}\zeta^{b\nu})] .
\ee
With the decomposition
$
B^a_{\mu\nu}=b^a_{\mu\nu}-\tilde\mathcal{F}^a_{\mu\nu}
$
and making use of Eq.~(\ref{YMEq}), we obtain,
\be
Z&=&\mathcal{Z}\int[db]\Det^{-\sfrac{1}{2}}(\mathbbm{b}+\mathbbm{F})
\exp(\I\sint d^4x\{
-\sfrac{1}{4}b^a_{\mu\nu}b^{a\mu\nu}
+
\nn
&&+\sfrac{1}{2}
(\mathcal{D}^{ac}_\kappa\tilde b^{c\kappa\mu})
[(\mathbbm{b}+\mathbbm{F})^{-1}]^{ab}_{\mu\nu}
(\mathcal{D}^{bd}_\lambda\tilde b^{d\lambda\nu})\}),
\label{gf:split}
\ee
where
$
\mathcal{Z}
=
\exp[\I\sint d^4x(
-\sfrac{1}{4}\mathcal{F}^a_{\mu\nu}\mathcal{F}^{a\mu\nu}
-\mathcal{A}^a_\mu J^{a\mu})] , 
$
$
\mathbbm{b}^{bc}_{\mu\nu}=g\tilde b^a_{\mu\nu}f^{abc} ,
$
and
$
\mathbbm{F}^{bc}_{\mu\nu}=g\mathcal{F}^a_{\mu\nu}f^{abc} .
$

Carrying out a gauge transformation $U$ of the background $J_\mu\rightarrow
UJ_\mu U^\dagger$ leads to $\mathcal{D}_\mu\rightarrow U\mathcal{D}_\mu
U^\dagger$ and $\mathcal{F}_{\mu\nu}\rightarrow
U\mathcal{F}_{\mu\nu}U^\dagger$. The gauge transformations $U$ which then 
appear in $Z$ can be removed by the same unitary transformation of the 
integration variable $b_{\mu\nu}\rightarrow Ub_{\mu\nu}U^\dagger$. Consistently,
$\mathcal{F}_{\mu\nu}+\tilde b_{\mu\nu}=B_{\mu\nu}\rightarrow
UB_{\mu\nu}U^\dagger$. $\mathcal{Z}$ is unaffected. Let us call these
type IB gauge transformations.

The generating functional is invariant as long as the
total $B_{\mu\nu}=b_{\mu\nu}-\tilde\mathcal{F}_{\mu\nu}$ transforms 
homogeneously.
This remains true especially for what one could call 
a type IIB transformation, where the background is left invariant and the 
fluctuation field accounts for the entire transformation, 
$
b_{\mu\nu}
\rightarrow 
U(b_{\mu\nu}-\tilde\mathcal{F}_{\mu\nu})U^\dagger+\tilde\mathcal{F}_{\mu\nu}
$.
After such a transformation, however, the transformed $b_{\mu\nu}^a$ field is in
general not a pure fluctuation field anymore; it obtains an expectation
value $-U\tilde\mathcal{F}_{\mu\nu}U^\dagger+\tilde\mathcal{F}_{\mu\nu}$. A
redecomposition into a true expectation value and true fluctuations would
reverse this transformation.

In the background field method \cite{BFM,DDD:BFM}, the gauge fixing term 
reads
$
-(\mathcal{D}^{ab}_\mu a^{b\mu})(\mathcal{D}^{ac}_\nu a^{c\nu})/(2\xi)
$
and is gauge invariant under type IA gauge transformations, 
$\mathcal{D}_\mu\rightarrow U\mathcal{D}_\mu U^\dagger$ and 
$a_\mu\rightarrow Ua_\mu U^\dagger$. Likewise, type IIA transformations
leave the background $\mathcal{A}^a_\mu$ invariant and 
$a_\mu\rightarrow UD_\mu U^\dagger-\mathcal{D}_\mu$. Again here, $a_\mu^a$ 
after the latter transformation has, in general, an expectation value 
and is consequently not a pure fluctuation field anymore. The aforementioned
redecomposition would undo the type IIA transformation.

In both cases the actions are manifestly gauge invariant under type I gauge
transformations. Type II transformations necessitate a redecomposition into
expectation value and fluctuations. In any case taking the contribution to a
quantity from the background and the
fluctuation field together admits finding a result invariant under both
types of gauge transformations. A main difference of the $B^a_{\mu\nu}$ with
respect to the $A^a_\mu$ field description is the absence of an explicit
gauge fixing term and consequently of ghost terms in the former.

The stationarity condition is derived by variation with respect to
$b^e_{\alpha\beta}$,
\be
0
=
-b^{e\alpha\beta}
-\mathcal{D}^{ae}_\kappa\epsilon^{\alpha\beta\kappa\mu}
[(\mathbbm{b}+\mathbbm{F})^{-1}]^{ab}_{\mu\nu}
(\mathcal{D}^{bd}_\lambda\tilde b^{\lambda\nu})
\nn
-\sfrac{1}{2}
(\mathcal{D}^{ac}_\kappa\tilde b^{c\kappa\mu})
[(\mathbbm{b}+\mathbbm{F})^{-1}]_{\mu\rho}^{af}
f^{efg}\epsilon^{\alpha\beta\rho\sigma}
\times\nn\times
[(\mathbbm{b}+\mathbbm{F})^{-1}]_{\sigma\nu}^{gb}
(\mathcal{D}^{bd}_\lambda\tilde b^{\lambda\nu})
.
\label{cleom}
\ee
[The determinant term does not contribute at this level as varying with 
respect to $\zeta^{b\nu}$ implies
$\zeta^{a\mu}(\mathbbm{b}+\mathbbm{F})_{\mu\nu}^{ab}=0$.]
Only if it has the solution $b^a_{\mu\nu}\equiv 0$ can $b^a_{\mu\nu}$ be
treated as pure fluctuation. Otherwise, the appropriate expansion point,
i.e., the correct vacuum, has to be determined by finding the solution of the 
previous equation. Remarkably, in that case, the expansion point would be 
different from $\mathcal{F}^a_{\mu\nu}$, the one in the vector field 
formulation of Yang--Mills theory. Situations where $\det\mathbbm{F}=0$, are
problematic in this respect because there Eq.~(\ref{cleom}) is ill-defined at 
$b^a_{\mu\nu}=0$. Among the settings belonging to this group is the trivial, 
i.e., background field free case. In this context this coincides with the 
observation that the zero field vacuum in Yang--Mills theories is unstable 
\cite{Nielsen:1978rm}. One may wonder, how the standard high-energy 
perturbative treatment comes about in the present formalism. There, at least
initially, Yang--Mills theory looks almost Abelian. Here, for 
$g\rightarrow 0$ (and without background) the Gaussian made up by the last 
term in the generating functional (\ref{gf:BF}) goes to a $\delta$ 
distribution and imposes several colour copies of the Maxwell equation as 
seen in the previous section in the Abelian case. \cite{DDD:ConstantFields}  

There are also nonzero configurations with $\det\mathbbm{F}=0$: 
$\mathbbm{F}$ is in the adoint representation. Hence, each
Lorentz component alone has zero eigenvalues. Therefore, to have
$\det\mathbbm{F}\neq 0$ one needs several Lorentz components whose
eigenvectors belonging to the zero eigenvalues are misaligned. 
Thus field configurations with a single Lorentz component have necessarily 
$\det\mathbbm{F}=0$. Among these are Coulomb fields, also those boosted onto
the light cone. Their application to the description of the initial
condition of heavy-ion collisions gives rise to instabilities \cite{McLV}. 

The customary generalisation of effective actions \cite{ConstantFieldHistory} 
in the presence of constant external field tensors to the non-Abelian case 
\cite{nabeffact} proceeds via covariantly constant fields, 
$
\mathcal{D}^{ab}_\lambda\mathcal{F}^b_{\mu\nu}=0~~\forall~~\lambda,\mu,\nu .
$
They are effectively quasi-Abelian and lead to a result analogous to the
Abelian. They also have $\det\mathbbm{F}=0$. This condition is also a
necessary condition for a Wu--Yang ambiguity \cite{WY} to appear in four
dimensions \cite{WY:Det4d}. A Wu--Yang ambiguous field tensor can be
realised by different gauge field configurations which are {\it not} gauge
equivalent. This implies that in such cases not all information about the
system or its background is contained in the field tensor. The covariant
derivative contains more information than its commutator. Thus,
one can also understand why the factor $\Det^{-1/2}\mathbbm{B}$
appears as Jacobian in the measure when translating the Yang--Mills
generating functional $Z$ from a vector to an antisymmetric tensor field
representation \cite{Deser:1976iy,B}: $B^a_{\mu\nu}$ is the conjugate of
$\mathcal{F}^a_{\mu\nu}$. Where the field tensor does not allow to
reconstruct the system uniquely, but the vector potential would, the
Jacobian becomes singular. One particular quantity which differs for
gauge inequivalent gauge field realisations for the same field
tensor is the Yang--Mills current $J^a_\mu$. In our case the full
information about the system is communicated from the $A^a_\mu$ to the
$B^a_{\mu\nu}$ representation through said $J^a_\mu$ and the (classical)
covariant derivative $\mathcal{D}_{\mu}^{ab}$. [See Eq.~(\ref{gf:BF}).]
A further conclusions \cite{Dietrich:2009..} is that in the presence of a 
Wu--Yang ambiguous background not all observables can be expressed in terms 
of invariants \cite{Invariants} constructed merely from
$\mathcal{F}^a_{\mu\nu}$. 


\vs

Let us take a look at an example where the classical field has
$\det\mathbbm{F}\neq0$ for which we would like to calculate the
effective action induced by the fluctuations. For tractability we choose a 
three-dimensional
Euclidean system with an $SU(2)$ gauge group. We start out with a generating
functional coupled to an external source just like at the beginning of this
section, separate off the fluctuations of the vector gauge field around the
background, and translate into a representation based on the variables 
$E_\mu^a$, the three-dimensional analogue of $B^a_{\mu\nu}$.
As counterpart to Eq.~(\ref{YMBF}) we find,
\be
Z\cong\int[da][dE]\exp[\sint d^3x(
-\sfrac{1}{2}E^a_\mu E^a_\mu
-\sfrac{1}{2}\tilde E^a_{\mu\nu} F^a_{\mu\nu}
-A^a_\mu J^a_\mu)] ,
\nonumber
\ee
where
$
\tilde E^a_{\mu\nu}=\I\epsilon_{\kappa\mu\nu}E^a_\kappa .
$
Integrating out the fluctuations yields,
\be
Z
&\cong&
\int[dE]\Det^{-\sfrac{1}{2}}\mathbbm{E}\exp\{\sint d^3x[
\nn
&&-\sfrac{1}{2}E^a_\mu E^a_\mu
-\sfrac{1}{2}\tilde E^a_{\mu\nu}\mathcal{F}^a_{\mu\nu}
-\mathcal{A}^a_\mu J^a_\mu
+\\
&&+\sfrac{1}{2}
(\mathcal{D}^{ac}_\kappa\tilde E^c_{\kappa\mu}-J^a_\mu)
(\mathbbm{E}^{-1})^{ab}_{\mu\nu}
(\mathcal{D}^{bd}_\lambda\tilde E^d_{\lambda\nu}-J^b_\nu)
]\} ,
\nonumber
\ee
where $\mathbbm{E}^{bc}_{\mu\nu}=g\tilde E^a_{\mu\nu}\epsilon^{abc}$.
The decomposition
$
\tilde E^a_{\mu\nu}
=
\tilde e^a_{\mu\nu}
+
\mathcal{F}^a_{\mu\nu}
$
gives,
\be
Z
&=&
\mathcal{Z}\int[de]\Det^{-\sfrac{1}{2}}(\mathbbm{e}+\mathbbm{F})
\exp(\sint d^3x\{
-\sfrac{1}{2}e^a_\mu e^a_\mu
+\nn
&&+\sfrac{1}{2}
(\mathcal{D}^{ac}_\kappa\tilde e^c_{\kappa\mu})
[(\mathbbm{e}+\mathbbm{F})^{-1}]^{ab}_{\mu\nu}
(\mathcal{D}^{bd}_\lambda\tilde e^d_{\lambda\nu})\}) .
\ee
Thus, in momentum space, for a constant background the fluctuation operator 
for $e^a_\mu$ reads,
\be
(G^{-1})^{bd}_{\alpha\beta}
=
\delta^{bd}\delta_{\alpha\beta}
-
(\delta^{ab}p_\kappa+\I g\epsilon^{afb}\mathcal{A}^f_\kappa)
\epsilon_{\alpha\kappa\mu}
\times\nn\times
(\mathbbm{F}^{-1})^{ac}_{\mu\nu}
(\delta^{cd}p_\lambda+\I g\epsilon^{ced}\mathcal{A}^e_\lambda)
\epsilon_{\nu\lambda\beta}.
\ee
Based on it we would like to calculate the effective Lagrangian,
\be
\mathcal{L}^{(1)}=\frac{1}{2}
\lim_{\varepsilon\rightarrow 0}
\int\frac{d^3p}{(2\pi)^3}
\ln\det 
\frac
{\mathbbm{F}G^{-1}(p)}
{(\varepsilon^2\mathbbm{F})G_\varepsilon^{-1}(p)} .
\ee
It is normalised with respect to the free part which is obtained from the expression in the full background field by the rescaling $\mathcal{A}^a_\mu\mapsto\varepsilon\mathcal{A}^a_\mu$ and taking $\varepsilon$ to zero at the end.
The factors of $\mathbbm{F}$ stem from the $\zeta^a_\mu$ integration.

For tractability we specialise to
$\mathcal{A}^1_1=\mathcal{A}^2_2=\mathcal{A}^3_3=\mathcal{A}$ and zero
otherwise, corresponding to
$\mathcal{F}^3_{12}=\mathcal{F}^1_{23}=\mathcal{F}^2_{31}=g\mathcal{AA}=\mathcal{F}$ 
and
$J_1^1=J_2^2=J_3^3=-2g^2\mathcal{AAA}$.
We find for the determinant,
\be
\det(G^{-1}G_\varepsilon/\varepsilon^2)
=
\varepsilon^{-10}
(p^4+|2g\mathcal{F}|^2)
(p^4+|2g\varepsilon^2\mathcal{F}|^2)^{-1}.
\nonumber
\ee
For $\mathcal{L}^{(1)}$ this leads to,
\be
\mathcal{L}^{(1)}
=
\lim_{\varepsilon\rightarrow 0}
\int\frac{d|p||p|^2}{(2\pi)^2}
\left(
\ln\frac{|p|^4+|2g\mathcal{F}|^2}{|p|^4}
-
\ln\varepsilon^{10}
\right).
\ee
The 
integral is IR finite and UV divergent (the last term).
After removing the divergent part (and taking the now trivial $\varepsilon\rightarrow0$ limit), $\mathcal{L}^{(1)}$ evaluates to,
\be
\mathcal{L}^{(1)}
=
|g\mathcal{F}|^\sfrac{3}{2}/(3\pi) .
\ee

The $\mathcal{F}$ dependent prefactor has its origin in the covariant
derivative and the momentum integral. When rescaling every momentum by
$|g\mathcal F|^{1/2}$ the measure picks up a factor of $|g\mathcal
F|^{d/2}$ in $d$ dimensions. The remaining integral is
field independent. In four dimensions this amounts to a factor
$\sim g^2\mathcal{F}_{\mu\nu}^a\mathcal{F}_{\mu\nu}^a$. Thus, a divergent
contribution from the integral can be handled by renormalisation.
The factor of $|g\mathcal{F}|^{d/2}$ is known from the strong field/massless 
limit of Abelian effective actions induced by scalars or fermions in constant 
fields \cite{Blau:1988iz}. 


\vs

{\it In conclusion}, we have analysed fluctuations of gauge field around a
classical configuration by means of a transformation from the
inhomogeneously transforming vector gauge field to a homogeneously
transforming antisymmetric tensor field.
In the Abelian case this procedure yields the same result as if one
integrated out $A_\mu$ in Landau gauge, with the difference that no gauge is
specified.

For non-Abelian fields the fluctuation analysis proceeds also without the
introduction of a gauge or ghosts. It leads to the
$\sim|g\mathcal{F}|^{d/2}$ behaviour of the effective action in $d$
dimensions. We checked this explicitly for a static background without
Wu--Yang ambiguity. For four dimensions this indicates a dependence $\sim
g^2\mathcal{F}_{\mu\nu}^a\mathcal{F}^{a\mu\nu}$ permitting the treatment of
infinite contributions by renormalisation.

Additionally, in the $B^a_{\mu\nu}$ field formulation, the criterion $\det\mathbbm{F}=0$ marks  background fields which give rise to instabilites, e.g., no field or a Coulomb field, which links them to Wu--Yang ambiguities.

It would be interesting to recast the present approach in the
framework of the worldline formalism \cite{WL}.


\vs

The author would like to acknowledge discussions with
G.~Dietrich, 
F.~Sannino,
K.~Socha,
M.~Svensson, 
A.~Swann,
and
U.~I.~S{\o}ndergaard.
The author's work was supported by the Danish Natural Science Research Council. 



\end{document}